\documentstyle[aps,preprint,eqsecnum,epsf]{revtex}

\global\firstfigfalse

\tighten
\draft

\begin{document}

\preprint{MAD-NT-96-03}
\title{Relations Between Fusion Cross Sections and Average Angular Momenta}
\author{A.B. Balantekin\thanks{E-mail address: {\tt
	baha@nucth.physics.wisc.edu}}
 and A.J. DeWeerd\thanks{E-mail address: {\tt
	deweerd@nucth.physics.wisc.edu}}}
\address{Department of Physics, University of Wisconsin,
        Madison, Wisconsin 53706 USA}
\author{S. Kuyucak\thanks{E-mail address: {\tt
	sek105@phys.anu.edu.au}}}
\address{Department of Theoretical Physics, Research School of Physical
	Sciences, \\
	Australian National University, Canberra, ACT 0200, Australia}

\date{June 21, 1996}
\maketitle
\vskip 0.2in
\begin{abstract}
We study the relations between moments of fusion cross sections and
averages of angular momentum. The role of the centrifugal barrier and
the target deformation in determining the effective barrier radius are
clarified. A simple method for extracting average angular momentum from fusion
cross sections is demonstrated using numerical examples as well as actual data.
\end{abstract}

\pacs{25.70.Jj, 21.60.Fw, 21.60.Ev}
\newpage


\section{Introduction}
The suggestion that the barrier distributions in subbarrier fusion
reactions could be determined directly from the cross section data
\cite{row91} has led to a renewed experimental activity in the field
\cite{wei91,lei93,lem93,mor94,ste95}.  Since barrier distributions are
proportional to $d^2(E\sigma)/dE^2$, very accurate measurements of
excitation functions at closely-spaced energies are required. Even with
excellent data, smooth barrier distributions can only be obtained under
certain model dependent assumptions \cite{izu95} (i.e. - well-chosen energy
spacing for calculating the second derivatives).  Recently, it has been
suggested that analyses using integrals over fusion data \cite{kra95}
provide model independent results, so these should be preferred over
analyses that rely on differentiation of data.  Specifically, the
incomplete $n$-th moments of $E\sigma$,
\begin{equation}
\label{nthmoment}
f_n(E) = n(n-1) \int_0^E dE' (E-E')^{n-2} E' \sigma(E'), \quad n\ge 2,
\end{equation}
were proposed as an alternative for comparing model calculations with
data. Unfortunately, these moments are not directly related to observables
so it is difficult to give a physical interpretation for them.  In
addition, one of the main advantages of using barrier distributions is
that they bring out important features in the data, but the the $n$-th moments
in Eq.~(\ref{nthmoment}) are even more featureless than the data from which
they are calculated.

The purpose of this paper is to point out how the average angular momentum,
which is a measurable quantity \cite{van92}, is related to moments of the
fusion cross section.  In order to turn these
relations into a practical tool to extract average angular
momentum directly from fusion data, we study the angular momentum and
deformation dependence of the effective barrier radius used in the
formalism.
In Section \ref{sec-general}, we review the relation between
cross section and average angular momentum and introduce an expression for
calculating the effective barrier radius. We derive improved relations in
Section \ref{sec-ldep} by using a better approximation for the transmission
probability for the $\ell$-th partial wave.  This provides a satisfactory
description of the relation between the cross section and average angular
momentum for a spherical target. In Section \ref{sec-def},
we present a qualitative discussion of deformation effects and propose a
a simple way to include it in the effective radius.
As a practical application of the method, we obtain the average angular
momentum from the fusion cross section data in the $^{16}$O+$^{154}$Sm system
and compare the results with the experiment. Finally, we draw conclusions from
this work.


\section{The General Method}
\label{sec-general}
The idea of expressing average angular momenta in terms of integrals over
functions of cross sections dates back to Ref.~\cite{bal86}. Due to absence
of sufficiently good subbarrier fusion data, the full potential of this
idea was not explored at that time.
To introduce the concepts involved, we first review the general relations
between moments of fusion cross sections and averages of powers of angular
momenta in a one-dimensional barrier penetration picture. Although this
is strictly valid only for spherical systems, it will provide
the basis for extension to deformed systems. In the usual partial wave
expansion, the total fusion  cross section at the center-of-mass energy
$E$ is written as
\begin{equation}
\label{sigma}
\sigma(E) = \sum_{\ell=0}^\infty \sigma_\ell(E) \, ,
\end{equation}
where the cross section for the $\ell$-th partial wave is
\begin{equation}
\label{sigmal}
\sigma_\ell(E) = \frac{\pi \hbar^2}{2\mu E}\, \left(2\ell+1 \right) T_\ell(E)
\,
\end{equation}
$T_\ell(E)$ the transmission probability for that partial wave, and $\mu$
the reduced mass.
For energies near the
Coulomb barrier, one can approximate the $\ell$-dependence in $T_\ell$ by using
the $s$-wave penetrability at a shifted energy \cite{bal83},
\begin{equation}
\label{tlappr}
T_\ell(E) = T_0 \left(E - \frac{\hbar^2 \ell(\ell+1)}{2\mu R^2(E)} \right) \, ,
\end{equation}
where $\mu R^2(E)$ is the effective moment of inertia of the system.  The
energy shift simply accounts for the change in the height of the barrier
resulting from the centrifugal potential.
Note that the effective radius is allowed to vary as a function of energy.
In Section~\ref{sec-ldep}, we will demonstrate how this approximation for
the transmission probability can be improved.  Substituting
Eqs. (\ref{sigmal}) and (\ref{tlappr}) into Eq.~(\ref{sigma}), converting the
sum over $\ell$ into an integral, and changing variables to
\begin{equation}
\label{eprime}
E' = E -  \frac{\hbar^2 \ell(\ell+1)}{2\mu R^2(E)} \, ,
\end{equation}
we obtain the expression
\begin{equation}
\label{rinv}
\sigma(E) = \frac{\pi R^2(E)}{E} \, \int_0^E dE' \, T_0(E') \,.
\end{equation}
We will use Eq.~(\ref{rinv}) to study the energy dependence
of the effective radius from numerical calculations.

The average angular momentum after fusion is assumed to be
\begin{equation}
\langle \ell \rangle = \frac{1}{\sigma(E)} \, \sum_{\ell=0}^\infty
\ell\sigma_\ell(E) \,.
\end{equation}
Following a procedure similar to the one used to obtain Eq.~(\ref{rinv}),
the average angular momentum can be written as
\begin{eqnarray}
\langle \ell \rangle = \frac{\pi R^2(E)}{E\sigma(E)} \int_0^E dE' \,
   T_0(E') \, \left\{\left[
      \frac{2\mu R^2(E)}{\hbar^2} \, (E-E') + \frac{1}{4} \right]^{1/2}
      - \frac{1}{2} \right\} \,.
\end{eqnarray}
Integrating by parts and using Eq.~(\ref{rinv}), we arrive at the desired
expression,
\begin{eqnarray}
\label{lint}
\langle \ell \rangle = \frac{\mu R^4(E)}{\hbar^2 E\sigma(E)} \int_0^E dE'
   \, \frac{E' \sigma(E')}{R^2(E')} \,
   \left[\frac{2\mu R^2(E)}{\hbar^2}\,(E-E')+ \frac{1}{4} \right]^{-1/2} ,
\end{eqnarray}
which relates the average angular momentum to a moment of the fusion cross
section.  Note that the factor of 2 in Eq.~(\ref{lint}) is missing in
Ref.~\cite{bal86}.  For practical applications of Eq.~(\ref{lint}),
it is important to note that, in order to obtain the detailed features of the
average angular momentum reliably, the cross section should be interpolated
rather than fit globally. We have found that a spline fit to the logarithm of
the cross section works quite well for this purpose.
This integral method with the assumption that the effective radius is
constant, has been applied to some nearly spherical systems \cite{das91}.
The results are in agreement with the data within the experimental errors
which, however, are rather large for a definitive confirmation that Eq.
(\ref{lint}) with a constant radius works.
The existing data for a variety of systems were also examined assuming a
constant effective radius\cite{bab93},
but that analysis used a fit of the the cross section to the exponential
of a polynomial so the features of the angular momenta are lost.

Higher moments of the angular momentum can be found by following similar
steps.  For example, the second moment of the angular momentum is given by
\begin{equation}
\label{l2int}
\langle \ell(\ell+1) \rangle =
   \frac{2\mu R^4(E)}{\hbar^2 E\sigma(E)} \, \int_0^E dE' \,
    \frac{E' \sigma(E')}{R^2(E')} \,.
\end{equation}
A similar expression for $\langle \ell^2 \rangle$ was given in
Ref.~\cite{das86}, but $\langle \ell \rangle$ was neglected
and  $R$ was assumed to be a constant. Under these assumptions,
$f_2$ in Eq.~(\ref{nthmoment}) can be related to $\langle \ell^2 \rangle$
through
\begin{equation}
\label{l2f2}
\langle \ell^2 \rangle \sim  \frac{\mu R^2}{\hbar^2 E\sigma} \, f_2(E).
\end{equation}
However, as has been pointed out previously \cite{bal86,row93}, taking $R$  to
be a constant is not a good approximation, especially for deformed nuclei.
Therefore, Eq.~(\ref{l2f2}) does not result in a reliable physical
interpretation for $f_2$.  There is little experimental data for higher moments
of $\ell$ due to the difficulty of these measurements, so we do not pursue them
here. The procedure for calculating them should be clear from the
preceding discussion.

\begin{figure}
\begin{center}
\epsfxsize=6.0in
\epsfbox{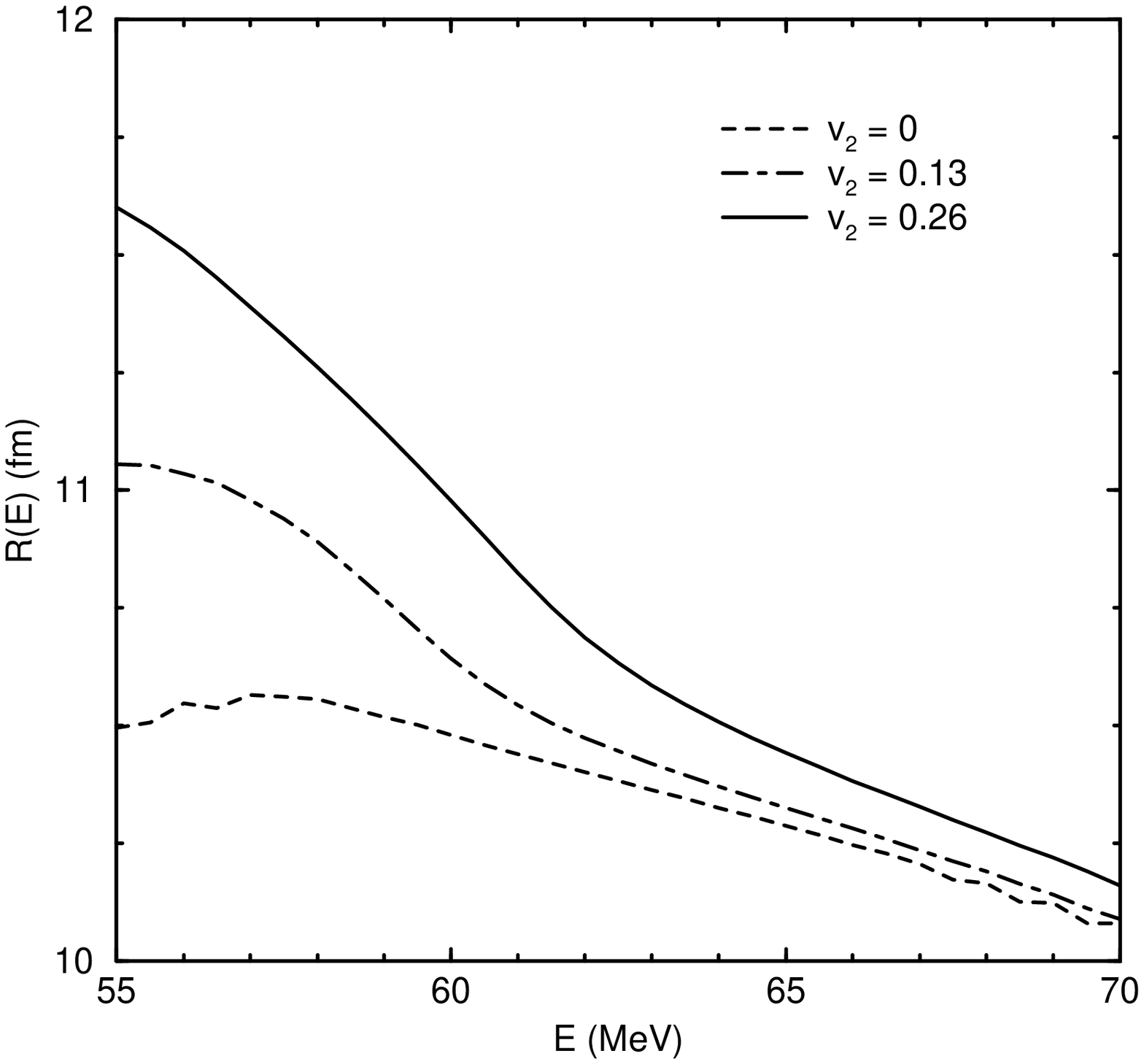}
\end{center}
\caption{The effective radius $R(E)$ extracted from fusion calculations for
the $^{16}$O+$^{154}$Sm system using Eq.~(\protect\ref{rinv}). The curves
correspond to spherical, vibrational and deformed nuclei with quadrupole
coupling strengths $v_2=0$, 0.13 and 0.26, respectively.}
\label{fig1}
\end{figure}

In order to put Eq.~(\ref{lint}) into use, we would like to understand the
origin of the energy dependence in the effective radius better.  For this
purpose, we use Eq.~(\ref{rinv}) as the defining relation for $R(E)$ and
study its deviation from a constant value.  We use values of $\sigma$ and
$T_0$ generated by the computer code IBMFUS \cite{ben93} which uses the
interacting boson model (IBM) \cite{iac87} to account for nuclear structure
effects and evaluates transmission probabilities numerically in the WKB
approximation.  For fusion reactions involving rare-earth nuclei, IBMFUS
has been shown to reproduce very well cross section and barrier
distribution systematics \cite{bal94a}, and average angular momentum data
\cite{bal94b}.  It is important to note that the centrifugal energy is
treated exactly in the WKB calculations (i.e. approximations such as
Eq.~(\ref{tlappr}) are not used).  In order to emphasize the effects of
nuclear structure, we kept the masses of the projectile ($^{16}$O) and
target ($^{154}$Sm) fixed and varied the quadrupole coupling parameter.
Fig.~\ref{fig1} shows the results obtained for $R(E)$ in three
cases corresponding to the target nucleus being spherical (no coupling),
vibrational (intermediate coupling) and deformed (strong coupling).  These
results demonstrate that the effective radius is not constant even for the
spherical case and deviates more as the coupling to structure
increases. The sharp increase in radius below the barrier with increasing
deformation is obviously due to selective sampling of the longer nuclear
axis.  The origin of the energy dependence of radius in the spherical
system is not that clear at this point, but the approximation used in
Eq.~(\ref{tlappr}) is an obvious suspect.  In view of the demonstrated
energy dependence of the effective radius, extracting the average angular
momentum from Eq.~(\ref{lint}) by assuming that $R(E)$ is constant will not
accurately predict $\langle \ell \rangle$ across a wide range of energies.

Attempts have been made to parameterize $R(E)$, but they have not been very
successful.  For example, a linear combination of the position of the barrier
peak $R_B$ and the Coulomb turning point
\begin{equation}
\label{rcoul}
R_C = Z_1 Z_2 e^2/E \,,
\end{equation}
has been suggested as a plausible choice \cite{bal83}
\begin{equation}
\label{rpar}
R(E) = \eta R_B + (1-\eta) R_C \,.
\end{equation}
In the spherical case, Eq.~(\ref{rpar}) provides a good description of
$R(E)$ in Fig.~\ref{fig1} with $\eta=0.78$.
However, is not clear how to include deformation effects in
Eq.~(\ref{rpar}) in a physically meaningful way.
Another expression for the effective radius, derived by assuming the nuclear
potential to be an exponential tail in the region of the barrier, is given by
\cite{row93}
\begin{equation}
R(E) = {1 \over 2} R_C \left[ 1 + \left(1-4a/R_C\right)^{1/2} \right] \,,
\end{equation}
where $a$ is the nuclear surface diffuseness.  This expression gives an
energy dependence which is too strong for values of $a$ in the range
0.6-1.2 fm.  Also it doesn't make any allowance for inclusion of
deformation effects, which are seen to have a significant influence on the
shape of $R(E)$.  Clearly, a better understanding of $R(E)$ is needed to
make further progress.


\section{An Improved Expression for the Penetrability}
\label{sec-ldep}

The prescription given in Eq.~(\ref{tlappr}) for approximating the $\ell$-wave
penetrability by the $s$-wave penetrability at a shifted energy utilizes only
the leading term in what is actually an infinite series expansion in
$\Lambda = \ell(\ell+1)$.  In this section, we derive the next term in this
expansion and show the resulting corrections to the calculations presented
in the Section \ref{sec-general}.  We also demonstrate how this can
explain the part of the energy dependence of the effective radius which
arises from the centrifugal potential.

The effect of the angular momentum on the penetrability is usually taken
into account by the shift it makes in the height of the
potential barrier.  The total potential for the $\ell$-th partial wave is
given by
\begin{equation}
\label{vsubl}
V_\ell(r) = V_N(r) + V_C(r) + \frac{\hbar^2 \ell(\ell+1)}{2\mu r^2} \,,
\end{equation}
where $V_N$ and $V_C$ are the nuclear and Coulomb potentials,
respectively.  Let $r_\ell$ denote the position of the peak of the $\ell$-wave
barrier which satisfies
\begin{equation}
\label{rldef}
\left. {\partial V_\ell(r) \over \partial r} \right|_{r=r_\ell} =0 \,,
\end{equation}
and
\begin{equation}
\left. {\partial^2 V_\ell(r) \over \partial r^2} \right|_{r=r_\ell} <0 \,,
\end{equation}
then the height of the barrier is given by $V_{Bl} = V_l(r_\ell)$.
We make the ansatz that the barrier position can be written as an infinite
series,
\begin{equation}
\label{rlans}
r_\ell = r_0 +  c_1 \Lambda + c_2 \Lambda^2 + \cdots,
\end{equation}
where the $c_i$ are constants.  Expanding all functions in Eq.~(\ref{rldef})
consistently in powers of $\Lambda$, we find that the first
coefficient is
\begin{equation}
\label{c1}
c_1 = - \, {\hbar^2 \over \mu\alpha r_0^3} \,,
\end{equation}
where $\alpha$ is the curvature of the $s$-wave barrier
\begin{equation}
\alpha= \left. - \, {\partial^2 V_0(r) \over \partial r^2 } \right|_{r=r_0}
\,.
\end{equation}
Substituting the leading order correction in the barrier position $r_\ell$
into Eq.~(\ref{vsubl}), we
find that to second order in $\Lambda$ the $\ell$-wave barrier height
is given by
\begin{equation}
\label{vbl}
V_{Bl} = V_{B0}
+ \frac{\hbar^2 \Lambda}{2\mu r_0^2}
+ \frac{\hbar^4 \Lambda^2}{2\mu^2 \alpha r_0^6} \,.
\end{equation}
Therefore, an improved approximation for the $\ell$-dependence in the
penetrability is given by
\begin{equation}
\label{tlcor}
T_\ell(E) = T_0 \left(E - \frac{\hbar^2 \Lambda}{2\mu r_0^2}
- -\frac{\hbar^4 \Lambda^2}{2\mu^2 \alpha r_0^6}\right).
\end{equation}
We give an alternative derivation of this expansion and discuss its
validity in the Appendix.

To examine the consequences of the improved expression for the penetrability,
we repeat the steps outlined in Section~\ref{sec-general} using
Eq.~(\ref{tlcor}) instead of Eq.~(\ref{tlappr}).
To the leading order in $1/\alpha$, this introduces the following correction
to Eq.~(\ref{rinv}),
\begin{equation}
\label{rcor}
\sigma(E) = \frac{\pi r_0^2}{E} \, \int_0^E dE' \, T_0(E')
\left[1-{4 \over \alpha r_0^2} (E-E')\right] \,.
\end{equation}
Comparing Eq.~(\ref{rinv}) with Eq.~(\ref{rcor}), we find that the
energy-dependent effective radius can be expressed as
\begin{equation}
\label{reff}
R^2(E) = r_0^2 \left[ 1 - {4 \over \alpha r_0^2} { \int_0^E dE' \, T_0(E')
(E-E') \over \int_0^E dE' \, T_0(E')} \right].
\end{equation}
This predicts the decrease in $R(E)$ as the energy increases that was shown
in Fig.~\ref{fig1}.
The calculation of the average angular momentum with
the modified equations introduces a similar leading order correction,
\begin{eqnarray}
\label{lcor}
\langle \ell \rangle = \frac{\mu r_0^2 }{\hbar^2 E\sigma(E)} \int_0^E dE'
   \, E' \sigma(E') \,
   \left[\frac{2\mu r_0^2}{\hbar^2}\,(E-E')+ \frac{1}{4} \right]^{-1/2}
   \left[1-\frac{7}{\alpha r_0^2} (E-E')\right]\,.
\end{eqnarray}
Since we have included a correction to $r_\ell$, this expression takes into
account the shifts in both the height and position of the barrier for
different partial waves.

It is easy to show that the expression for effective radius given in
Eq.~(\ref{reff}) is consistent with the result obtained by weighted averaging
over the barrier positions of all of the partial waves using
Eq.~(\ref{rlans}),
\begin{equation}
\label{rav1}
\langle r_\ell \rangle
= {\sum_\ell \sigma_\ell r_\ell \over \sum_\ell \sigma_\ell}
= r_0 - {\hbar^2 \over 2\mu \alpha r_0^3} \langle \Lambda \rangle + \cdots.
\end{equation}
To a first approximation, $\langle \Lambda \rangle$ is given by
Eq.~(\ref{l2int}) with $R(E)=r_0$. Substituting Eq.~(\ref{rinv}) for
$E'\sigma(E')$, Eq.~(\ref{rav1}) becomes
\begin{equation}
\label{rav2}
\langle r_\ell \rangle =
r_0 - {2\pi r_0 \over \alpha E\sigma} \int_0^E dE' \int_0^{E'}dE'' T_0(E'').
\end{equation}
After integration by parts, Eq.~(\ref{rav2}) becomes
\begin{equation}
\label{rav3}
\langle r_\ell \rangle =
r_0 - {2\pi r_0 \over \alpha E\sigma} \int_0^E dE' T_0(E') (E-E').
\end{equation}
Substituting Eq.~(\ref{rinv}) with $R(E)=r_0$ for $\sigma$ and squaring the
result, we recover an expression that is consistent with Eq.~(\ref{reff}) to
the first order in $1/\alpha$.

The effect of the correction to the barrier position on the distribution of
barriers is straightforward in the form given by Ackermann\cite{ack95},
which is in terms of first derivatives of the angular momentum
distribution.  This expression can be written as
\begin{equation}
\label{bardist}
D(E') = \frac{4\mu^2 R^2 E}{(2\ell+1)^2 \pi\hbar^2} \,
\frac{d\sigma_\ell(E)}{d\ell} \,,
\end{equation}
where the shifted energy is
\begin{equation}
E' = E - \frac{\hbar^2 \ell(\ell+1)}{2\mu R^2} \,.
\end{equation}
Knowledge of the angular momentum distribution at an energy $E$ allows
the calculation of the barrier distribution between
$E- \hbar^2 \ell_{max}(\ell_{max}+1) / 2\mu R^2$ and $E$, where $\ell_{max}$ is
the largest angular momentum for which the partial cross section is measured.
The first order correction to Eq.~(\ref{bardist}) is obtained by
substituting the position of the $\ell$-wave barrier for the effective radius
of
the associated partial-wave cross section
\begin{equation}
R \rightarrow r_\ell \approx r_0 + c_1 \Lambda \,.
\end{equation}
When the angular momentum distribution is determined at 65 MeV for the
$^{16}$O+$^{154}$Sm system, the correction to $D(E')$ is about 12\% at 55
MeV and it will be less for higher energies.  The other errors involved in
calculating the barrier distribution are typically larger than that, so
this correction can be safely ignored.

\begin{figure}
\begin{center}
\epsfxsize=6.0in
\epsfbox{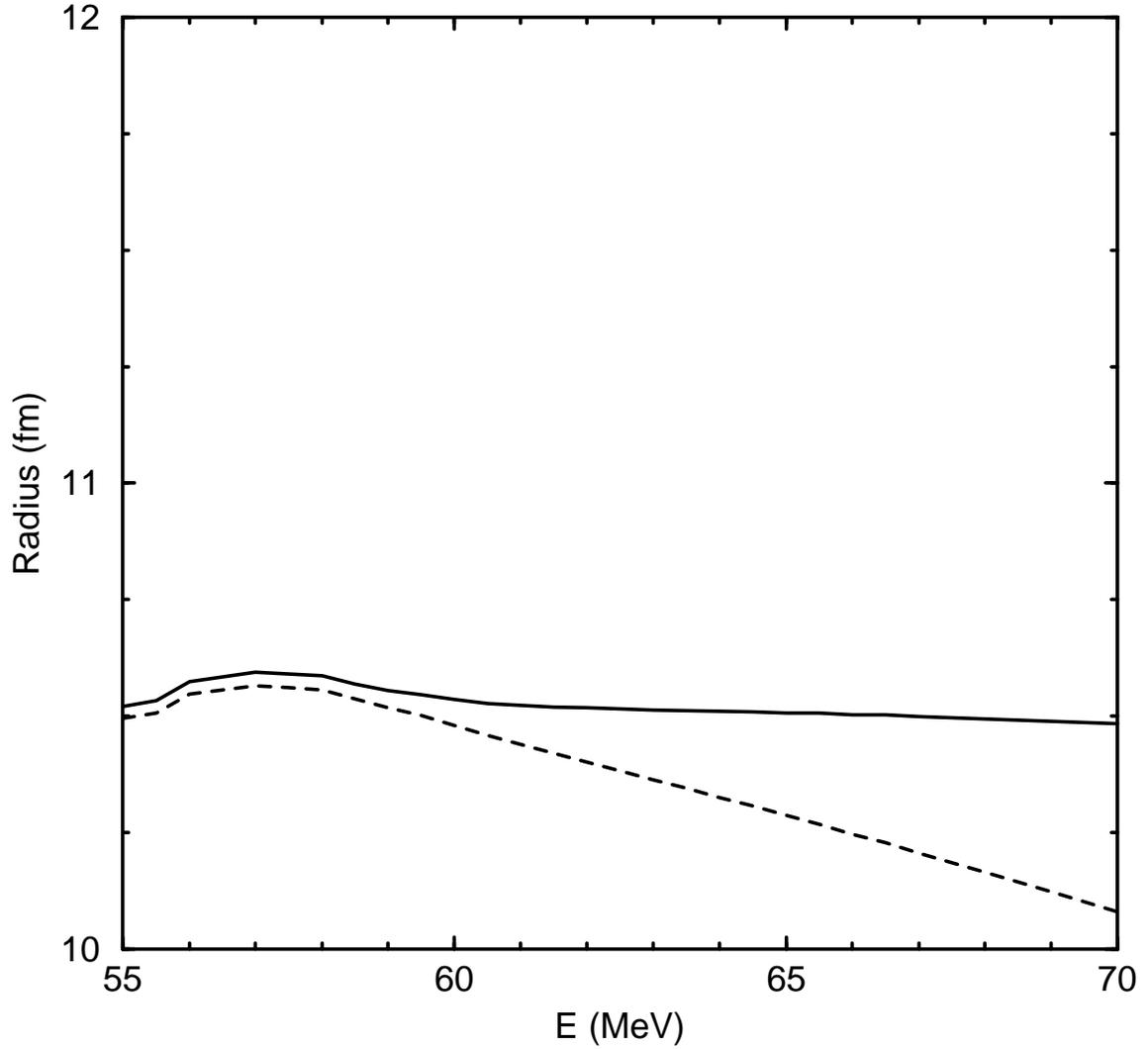}
\end{center}
\caption{The $s$-wave barrier position $r_0$ (solid curve) extracted from
fusion calculations for the $^{16}$O+$^{154}$Sm system with no coupling
using Eq.~(\protect\ref{rcor}).  For comparison, the values of $R(E)$
(dashed curve) calculated for the same reaction using
Eq.~(\protect\ref{rinv}) are also shown.}
\label{fig2}
\end{figure}

In order to check whether or not the correction in Eq.~(\ref{tlcor}) is
sufficient to account for the shift in the peak of the barrier due to the
angular momentum of the system, we calculated $r_0$ as a function of energy
for a spherical target using Eq.~(\ref{rcor}).  The same numerical
values of $\sigma(E)$ and $T_0(E)$ obtained from IBMFUS as in
Fig.~\ref{fig1} and the known value of $\alpha$ for the potential
barrier were used in this calculation. The results for the $s$-wave barrier
radius $r_0$ is shown in Fig.~\ref{fig2} and the effective radius
$R(E)$, extracted using Eq.~(\ref{rinv}), is shown for comparison.  The
results for $r_0$ are nearly independent of energy as we expect.

We would like to extract $r_0$ and $\alpha$ directly from cross section data.
Eq.~(\ref{rcor}) can be rewritten as
\begin{equation}
E \sigma(E) = \pi r_0^2 \, \int_0^E dE' \, T_0(E')
- -{4 \pi \over \alpha}\,E \int_0^E dE' \, T_0(E')
+{4 \pi \over \alpha} \int_0^E E' dE' \, T_0(E') \,.
\end{equation}
Using partial integration, the two integrals can be expressed as
\begin{equation}
\label{tint1}
\int_0^E dE' \, T_0(E') = E T_0(E)
- -\int_0^E E' dE' \, {dT_0 \over dE'} \,,
\end{equation}
and
\begin{equation}
\label{tint2}
\int_0^E E' dE' \, T_0(E') =  {E^2 T_0(E) \over 2}
- - \int_0^E {{E'}^2 \over 2} dE' \, {dT_0 \over dE'} \,.
\end{equation}
For $E \gg V_{B0}$, the right hand sides of
Eqs. (\ref{tint1}) and (\ref{tint2}) become $E-q_1$ and ${E^2 / 2} - q_2$,
respectively, where
\begin{equation}
\label{q1def}
q_1 = \int_0^E E' dE' \, {dT_0 \over dE'} \,,
\end{equation}
and
\begin{equation}
\label{q2def}
q_2 = {1\over 2}\int_0^E {E'^2} dE' \, {dT_0 \over dE'} \,.
\end{equation}
For energies above the barrier, $dT_0/dE'$ goes to 0, so $q_1$ and $q_2$
become constants.
Therefore, for high energies $E\sigma(E)$ is a quadratic in $E$,
\begin{equation}
E \sigma(E) = - \left({2 \pi \over \alpha} \right) E^2
+ \left(\pi r_0^2 + {4 \pi \over \alpha} \, q_1 \right) E
- - \left(\pi r_0^2 q_1 + {4 \pi \over \alpha} \, q_2\right) \,.
\end{equation}
Classically,
\begin{equation}
{dT_0 \over dE} = \delta(E - V_{B0}) \,,
\end{equation}
where $V_{B0}$ is the barrier height, so $q_1 = V_{B0}$ and $q_2 =
V_{B0}^2/2$.  Quantum mechanically, these expressions for $q_1$ and $q_2$
are also approximately true for energies above the barrier.  Using the
values of $T_0$ generated from the code IBMFUS \cite{ben93} in
Eqs. (\ref{q1def}) and (\ref{q2def}), we have found that the error in these
approximations at an energy of 70 MeV are both less than $0.05\%$ for an
O+Sm system with no coupling where $V_{B0}\approx 59 {\rm MeV}$.
Therefore, the product of the cross section and the energy can be fit
at high energies with the expression
\begin{eqnarray}
\label{esfit}
E \sigma(E)
&=& - \left({2 \pi \over \alpha} \right) E^2
+ \left(\pi r_0^2 + {4 \pi V_{B0} \over \alpha}  \right) E
- - \left(\pi r_0^2 V_{B0} + {2 \pi V_{B0}^2 \over \alpha} \right) \nonumber\\
&=& \pi r_0^2 \left( E-V_{B0} \right) - {2 \pi \over \alpha} \left( E-V_{B0}
\right)^2 \,.
\end{eqnarray}
in order to determine $\alpha$ and $r_0$.  Of course, this requires high
precision fusion data for energies above the barrier which may be difficult to
obtain due to the competing processes.

\begin{figure}
\begin{center}
\epsfxsize=6.0in
\epsfbox{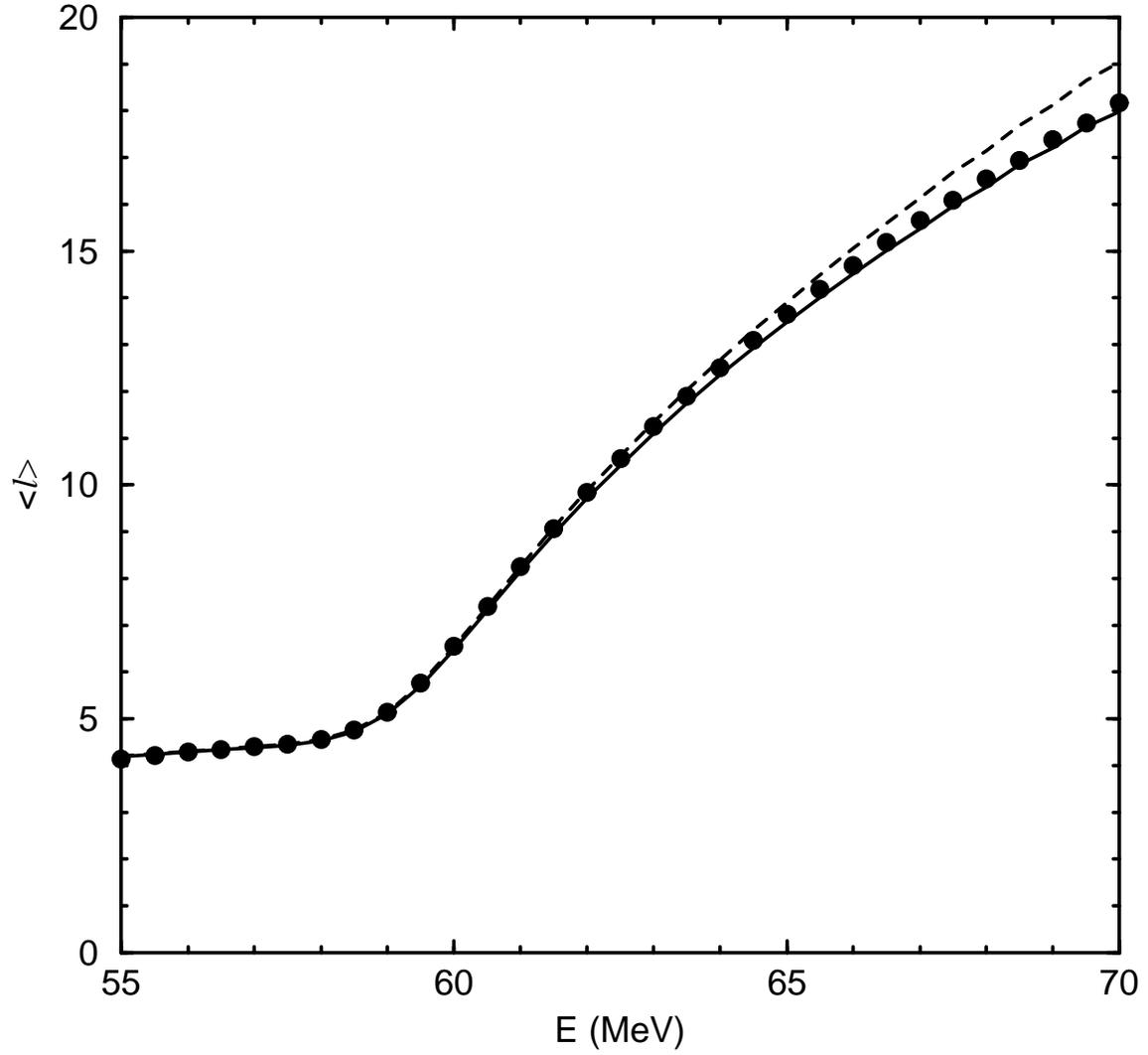}
\end{center}
\caption{ Comparison of average angular momenta calculated using
Eq.~(\protect\ref{lcor}) (solid curve) and Eq.~(\protect\ref{lint}) (dashed
curve) to the values calculated by IBMFUS (points) for the
$^{16}$O+$^{154}$Sm system with no coupling.}
\label{fig3}
\end{figure}

As a test of the formalism, we apply the results derived in this section to
the fusion cross section ``data" generated by IBMFUS for a the system of
$^{16}$O + $^{154}$Sm, where both nuclei are taken to be spherical. First,
$\alpha$ and $r_0$ are determined from Eq.~(\ref{esfit}) by a fit to the
$\sigma$ ``data". Then these values are
employed in Eq.~(\ref{lcor}) and the average angular momenta are extracted
from the $\sigma$ ``data" through numerical integration.
Fig.~\ref{fig3} shows a comparison of the average angular momenta
calculated using Eq.~(\ref{lcor}) with those obtained from IBMFUS directly.
The agreement is very good at all energies.
For reference, results obtained from Eq.~(\ref{lint}) assuming a constant
radius are also shown (dashed line). As expected, the modified expression
(\ref{lcor}) leads to a clear improvement at high energies where the effective
radius varies most (cf. Fig.~\ref{fig2}). This example gives confidence
that one can use Eq.~(\ref{lcor}) in extracting average angular
momenta directly from the fusion data for spherical systems.


\section{Target Deformation Effects}
\label{sec-def}

Now that we have an improved description of the effects due to the
centrifugal barriers, we consider the effects of target deformation.
The shape of an axially symmetric deformed nucleus can be described by
\begin{equation}
\label{Rtdeform}
{\cal R}_t(\theta) = {\cal R}_{t0} \left( 1 + \beta Y_{20}(\cos \theta)\right)
\,.
\end{equation}
where ${\cal R}_{t0}$ is the radius for an undeformed target.
In order to qualitatively study deformation effects, we approximate the target
with a simple two-level system which displays the important features. As shown
in Ref. \cite{nag86}, fusion of a deformed nucleus with finite number of levels
($n$) can be described by sampling $n$ orientations of Eq.~(\ref{Rtdeform})
with their respective weights. For a two-level system, the orientations
$\theta_1=70.12^\circ$ and $\theta_2=30.55^\circ$ contribute with the weight
factors $w_1=0.652$ and $w_2=0.348$, respectively.

To proceed, we need to find out how the barrier position and height
changes with orientation, which can be calculated from the total potential
in a straightforward manner.
For the nuclear potential, we use the usual Woods-Saxon form with the target
radius given by Eq.~(\ref{Rtdeform})
\begin{equation}
V_N(r,\theta)=-V_{N0} \left[ 1 + \exp\left( {r-{\cal R}_p-{\cal
R}_t(\theta)}\over a \right)
\right]^{-1} \,.
\end{equation}
The Coulomb potential can be calculated from a multipole expansion and,
to leading order in $\beta$, is given by
\begin{equation}
\label{vcoul}
V_C(r,\theta)={Z_1 Z_2 e^2 \over r}
\left(1 + {3\over 5} {\beta {\cal R}_{t0}^2 Y_{20}(\cos \theta) \over
r^2}\right)
= {A \over r} \left(1 + {3\over 5} {{\cal R}_{t0}({\cal R}_t(\theta) - {\cal
R}_{t0}) \over r^2}\right) \,,
\end{equation}
where $A=Z_1 Z_2 e^2$.
Using the notation of Section \ref{sec-ldep}, the equation
for finding the peak of the $s$-wave barrier is
\begin{equation}
\label{defr0}
\left.
- {A \over r^2}
- {9\over 5} {A {\cal R}_{t0}({\cal R}_t - {\cal R}_{t0}) \over r^4}
+ {V_{N0} \over a}
  {\exp\left(\left(r-{\cal R}_p-{\cal R}_t\right)/ a \right) \over
   \left[ 1+\exp\left(\left({r-{\cal R}_p-{\cal R}_t}\right) /a
\right)\right]^2} \,
 \right|_{r=r_0}
 = 0 \,,
\end{equation}
and the height of the $s$-wave barrier is
\begin{equation}
\label{defVB0}
V_{B0} = {A\over r_0}
\left(1 + {3\over 5} {{\cal R}_{t0}({\cal R}_t - {\cal R}_{t0}) \over
r_0^2}\right)
  - V_{N0} \left[ 1+\exp\left({r_0 -{\cal R}_p - {\cal R}_t}\over a
\right)\right]^{-1} .
\end{equation}
The $\theta$-dependence is suppressed in the above equations for convenience,
but both $r_0$ and $V_{B0}$ depend on the target orientation.

\begin{figure}
\begin{center}
\epsfxsize=6.0in
\epsfbox{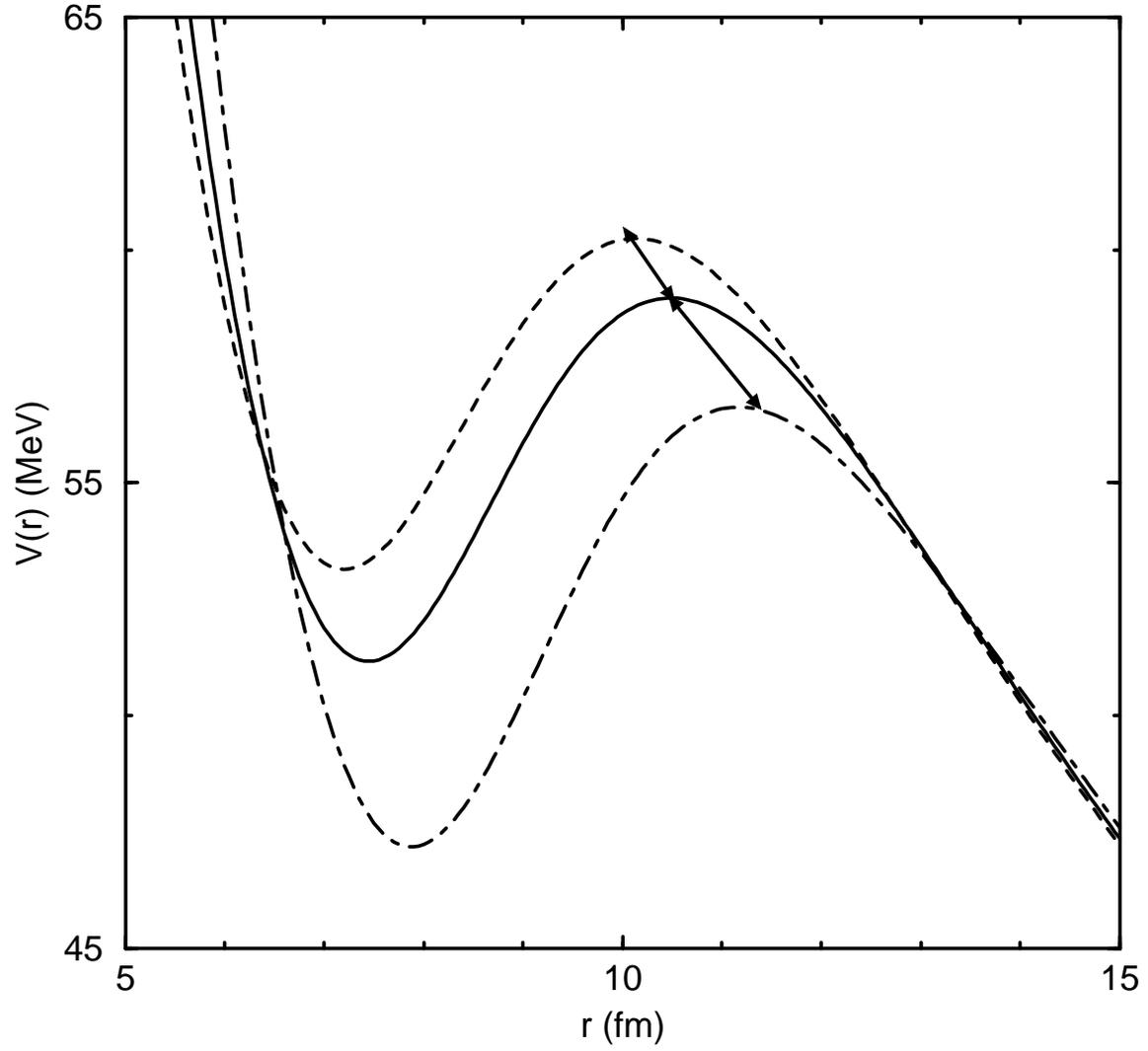}
\end{center}
\caption{The solid curve is the total potential for the $^{16}$O+$^{154}$Sm
system when the target is taken to be spherical.  The dashed
($\lambda=-0.327$) and dot-dashed ($\lambda=0.613$) curves are the potentials
for two-level approximation with $\beta=0.25$.  The arrows show the shifts
predicted for the barrier peaks by Eqs. (\protect\ref{deltav}) and
(\protect\ref{deltar}).}
\label{fig4}
\end{figure}

The rate of change in the barrier height due to the deformation is given by
\begin{equation}
\label{dvdRt}
{dV_{B0} \over d{\cal R}_t} =
 {dV_{B0} \over dr_0}{dr_0 \over d{\cal R}_t}
+ {3 A {\cal R}_{t0} \over 5 r_0^3}
 -{V_{N0} \over a}
  {\exp\left(\left({r_0-{\cal R}_p-{\cal R}_t}\right)/ a \right) \over
   \left[ 1+\exp\left(\left(r_0-{\cal R}_p-{\cal R}_t\right) /a
\right)\right]^2} \,.
\end{equation}
In Eq.~(\ref{dvdRt}), $dV_{B0}/dr_0=0$ by definition and using
Eq.~(\ref{defr0}), the second term can be simplified to give
\begin{equation}
\left. {dV_{B0} \over d{\cal R}_t} \right|_{{\cal R}_t={\cal R}_{t0}} = -\,{A
\over r_0^2}
+ {3 A {\cal R}_{t0} \over 5 r_0^3} \,.
\end{equation}
To find a similar expression for the barrier position, we differentiate
Eq.~(\ref{defr0}) with respect to ${\cal R}_t$
\begin{eqnarray}
\lefteqn{
\left[ {2A \over r_0^3}+{36 A\over 5} {{\cal R}_{t0}({\cal R}_t - {\cal
R}_{t0})\over r_0^5}\right]
   {dr_0 \over d{\cal R}_t}
 + {9 A {\cal R}_{t0} \over 5 r_0^4}
 + \left({dr_0 \over d{\cal R}_t} -1 \right) } \nonumber\\[0.25cm]
& &\times {V_{N0} \over a^2} {\exp\left(\left({r_0-{\cal R}_p-{\cal
R}_t}\right)/a\right) \over
   \left[ 1+ \exp\left(\left({r_0-{\cal R}_p-{\cal R}_t}\right)/ a
\right)\right]^2}
   \left[ 1 - {2 \exp\left(\left({r_0-{\cal R}_p-{\cal R}_t}\right)/ a \right)
\over
     \left[ 1+ \exp\left(\left({r_0-{\cal R}_p-{\cal R}_t}\right)/ a
\right)\right]} \right]
 = 0 \,.
\end{eqnarray}
Using Eq.~(\ref{defr0}), we can solve for the rate of change in
the $s$-wave barrier position due to the change in the target radius
\begin{equation}
\left. {dr_0 \over d{\cal R}_t} \right|_{{\cal R}_t={\cal R}_{t0}} = \frac
{ Q - {9a {\cal R}_{t0} / 5r_0^2} }
{{Q + {2a / r_0}  }} \,,
\end{equation}
where
\begin{equation}
Q = {2\over V_{N0}} \left({A\over r_0} - V_{B0} \right) -1 \,.
\end{equation}
For a given orientation $\theta$, the shift in the $s$-wave barrier height is
given approximately by
\begin{equation}
\label{deltav}
\delta V_{B0} = \left.{dV_{B0} \over d{\cal R}_t} \right|_{{\cal R}_t={\cal
R}_{t0}} \delta {\cal R}_t
 = \left( -{A \over r_0^2}  + {3 A {\cal R}_{t0} \over 5 r_0^3} \right)
   \sqrt{5 \over 4\pi} \, \beta {\cal R}_{t0} P_l(\cos \theta) \,,
\end{equation}
and, similarly, the shift in the $s$-wave barrier position is approximately
\begin{equation}
\label{deltar}
\delta r_0 = \left. {dr_0 \over d{\cal R}_t} \right|_{{\cal R}_t={\cal R}_{t0}}
\delta {\cal R}_t
  =  \left. {dr_0 \over d{\cal R}_t}\right|_{{\cal R}_t={\cal R}_{t0}} \sqrt{5
\over 4\pi} \,
     \beta {\cal R}_{t0} P_l(\cos\theta) \,.
\end{equation}
These expressions account for the changes due to deformation fairly accurately
as can be seen in Fig.~\ref{fig4}.

\begin{figure}
\begin{center}
\epsfxsize=6.0in
\epsfbox{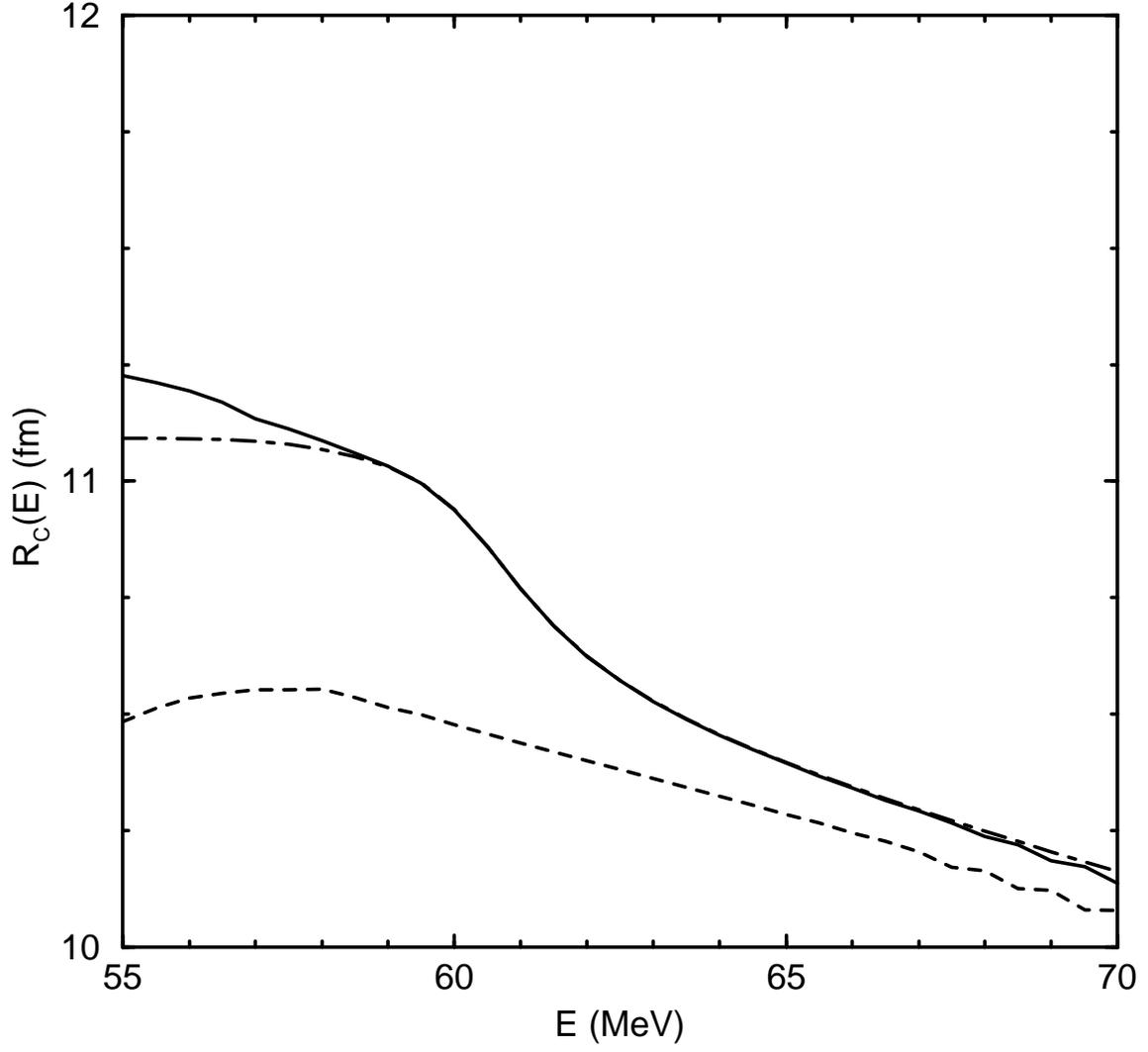}
\end{center}
\caption{The effective radius from a two-level calculation for the
$^{16}$O+$^{154}$Sm system with $\beta=0.25$.  The solid curve is
calculated using the definition, Eq.~(\protect\ref{rcdef}), and the dot-dashed
curve using Eq.~(\protect\ref{rcoupl}), which assumes $\alpha$
is constant. For comparison, the dashed curve is the result for a spherical
target.}
\label{fig5}
\end{figure}

The total fusion cross section for a two-level system is given by \cite{nag86}
\begin{equation}
\sigma_T(E) = w_1 \sigma(E,\lambda_1) + w_2 \sigma(E,\lambda_2) \,,
\end{equation}
where $\sigma(E,\lambda_i)$ is the cross section for the $i$-th
orientation and $\lambda_i=P_2(\cos \theta_i)$.  To simplify the notation,
we introduce
\begin{equation}
F=\sqrt{5 \over 4\pi}\, \left. {dr_0 \over d{\cal R}_t} \right|_{{\cal R}_t
 ={\cal R}_{t0}} {\cal R}_{t0} \,,
\end{equation}
so that, for a given orientation, the peak of the $s$-wave barrier in
Eq.~(\ref{rcor}) is replaced by
\begin{equation}
r_0 \rightarrow r_0 +  F \beta \lambda_i\,.
\end{equation}
After this substitution, the cross section for each level
becomes
\begin{eqnarray}
\sigma(E,\lambda_i) &=& \frac{\pi}{E} \, \int_0^E dE' \, \left\{
 r_0^2 T_0(E',\lambda_i)
   + F \beta r_0  \lambda_i T_0(E',\lambda_i)
   + F^2 \beta^2 \lambda_i^2 T_0(E',\lambda_i)  \right\} \nonumber\\[0.25cm]
&&- \frac{4\pi}{\alpha E} \, \int_0^E dE' \, T_0(E',\lambda_i) (E-E')  \,,
\end{eqnarray}
where $T_0(E,\lambda_i)$ is the transmission probability for $i$-th
orientation.
The curvature of the barrier, $\alpha$, is expected to have a second order
dependence on $\beta$, hence it is assumed to be constant in this leading order
calculation.
Defining the coupled $s$-wave transmission probability as
\begin{equation}
T_0^C(E) = w_1 T_0(E,\lambda_1) + w_2 T_0(E,\lambda_2) \,,
\end{equation}
the total cross section becomes
\begin{eqnarray}
\label{sigmacoupl}
\sigma_T(E) &=& \frac{\pi}{E} \, \left\{ r_0^2
 + 2F \beta r_0 \, \frac
  {\int_0^E dE' \, \left[ w_1 \lambda_1 T_0(E',\lambda_1)
   + w_2 \lambda_2 T_0(E',\lambda_2) \right]}
  {\int_0^E dE' \, T_0^C(E')}  \right.  \nonumber\\[0.25cm]
  & & \hspace{1cm} \left. + F^2 \beta^2 \, \frac
  {\int_0^E dE' \,\left[w_1 \lambda_1^2 T_0(E',\lambda_1)
   + w_2 \lambda_2^2 T_0(E',\lambda_2) \right] }
  {\int_0^E dE' \, T_0^C(E')} \right\}
{\int_0^E dE' \, T_0^C(E')} \nonumber\\[0.25cm]
& & -{4\pi \over \alpha E} \, \int_0^E dE' \, T_0^C(E') (E-E') \,.
\end{eqnarray}
The cross section can be written in the form of Eq.~(\ref{rinv}) as
\begin{equation}
\label{rcdef}
\sigma_T(E) = \frac{\pi}{E} \, R_C^2(E) \int_0^E dE' \, T_0^C(E') \,,
\end{equation}
so that the effective radius with coupling is
\begin{eqnarray}
\label{rcoupl}
R_C^2(E) &=& r_0^2
 + 2F \beta r_0 \, \frac
  {\int_0^E dE' \, \left[ w_1 \lambda_1 T_0(E',\lambda_1)
   + w_2 \lambda_2 T_0(E',\lambda_2) \right]}
  {\int_0^E dE' \, T_0^C(E')} \nonumber\\[0.25cm]
 & & + F^2 \beta^2 \, \frac
  {\int_0^E dE' \,\left[w_1 \lambda_1^2 T_0(E',\lambda_1)
   + w_2 \lambda_2^2 T_0(E',\lambda_2) \right] }
  {\int_0^E dE' \, T_0^C(E')} \nonumber\\[0.25cm]
 & & - {4 \over \alpha} \, \frac {\int_0^E dE' \, T_0^C(E') (E-E')}
  {\int_0^E dE' \, T_0^C(E')} \,.
\end{eqnarray}
Since $w_1 \lambda_1 = w_2 \lambda_2$ and both $T_0(E',\lambda_1)$ and
$T_0(E',\lambda_2)$ approach one very quickly for energies above the barriers,
the second term in Eq.~(\ref{rcoupl}) becomes zero for high energies.  On
the other hand, the third term of that equation is always positive, so by
comparison with Eq.~(\ref{reff}) it is easy to see that the effective radius
is slightly higher at large energies for the deformed case than for the
uncoupled case.  This argument also holds for multi-level systems, since
\begin{equation}
\int P_l(\cos \theta) \, d(\cos \theta) = 0 \,.
\end{equation}
The effective radius predicted by Eq.~(\ref{rcoupl}) (dashed curve) and the
result from definition of Eq.~(\ref{rcdef}) (solid curve) are shown in
Fig.~\ref{fig5}.  That the two curves are in good agreement is an
indication that the curvature is approximately unchanged as assumed in
Eq.~(\ref{rcoupl}).  This simple model exhibits the main features seen in
Fig.~\ref{fig1}; at low energies the difference between the
deformed and spherical cases becomes larger and this effect increases with
the deformation.

\begin{figure}
\begin{center}
\epsfxsize=6.0in
\epsfbox{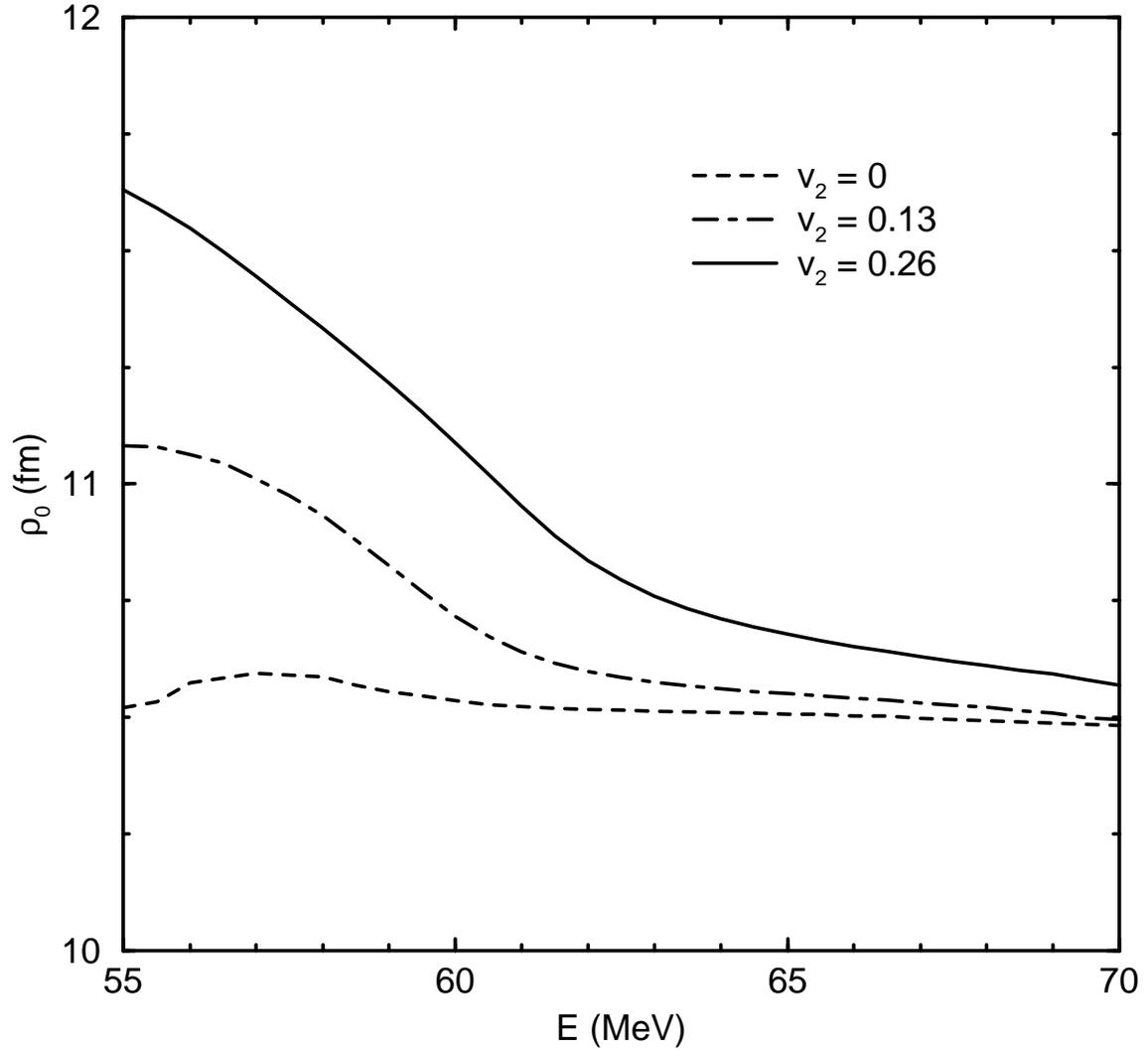}
\end{center}
\caption{$\rho_0$ extracted from fusion
calculations using Eq.~(\protect\ref{rhodef}) for the $^{16}$O+$^{154}$Sm
system with with quadrupole coupling strengths $v_2=0$, 0.13 and 0.26,
respectively.}
\label{fig6}
\end{figure}

The preceding argument provides an explanation of how the effective radius
varies with the energy. Unfortunately, it is not easy to incorporate the
deformation effects in the calculation of average angular momentum as was
done for the centrifugal barrier in Section \ref{sec-ldep}. To make
progress, we take
a phenomenological approach and introduce the quantity
\begin{equation}
\label{rhodef}
\rho_0^2(E) = \frac {E\sigma/\pi + (4/\alpha) \int_0^E dE' \, T_0(E') (E-E')}
  {\int_0^E dE' \, T_0(E')},
\label{rdef}
\end{equation}
For a single barrier, $\rho_0(E)$ is simply the location of the $s$-wave
barrier $r_0$ (cf. Eq.~(\ref{rcor})), so it is actually energy independent.
When couplings to target deformation are introduced, there is a
distribution of barriers.  In this case, $\rho^2_0(E)$ is a suitable
average of the location of the barrier peaks.  Using the
values of $\sigma$ and  $T_0$ generated by IBMFUS in Eq.~(\ref{rdef}) again, we
calculate $\rho_0(E)$ for the three quadrupole coupling strengths used
in the previous section.  In contrast to the effective
radii in Fig.~\ref{fig1}, the results shown in Fig.~\ref{fig6} level off at high
energies.  They can be parametrized using
a simple Fermi function
\begin{equation}
\label{r0ofe}
\rho_0(E) = r_0 + \frac{\delta}{1+\exp((E-V_{B0})/W)}.
\end{equation}
Here $r_0$ is the asymptotic value at large $E$ and $V_{B0}$ is the barrier
height, which are determined from the fusion data using Eq.~(\ref{esfit}).
$r_0+\delta$ corresponds to the asymptotic value at low energies and can be
calculated from Eq.~(\ref{Rtdeform}) at $\theta=0$. The only quantity in
Eq.~(\ref{r0ofe}) that is not determined from data is the width $W$.
The fits to the curves in  Fig.~\ref{fig6} results in values around
$2\pm0.3$ for $W$ with a mild dependence on $\beta$ (increases with $\beta$).
Since the precise value of $W$ makes no tangible difference, the slight
uncertainty in its value is not important for the purposes of this paper.

\begin{figure}
\begin{center}
\epsfxsize=6.0in
\epsfbox{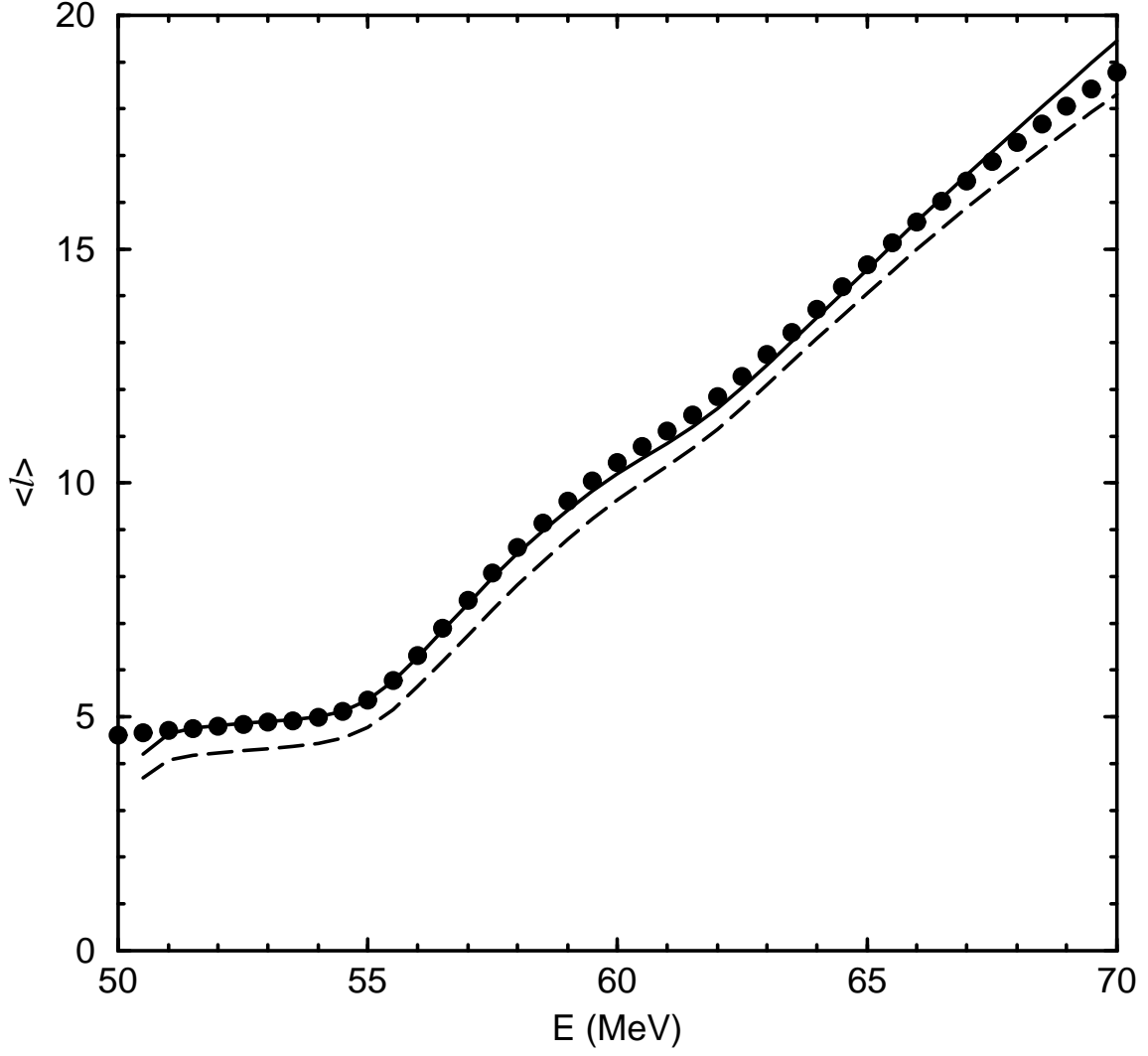}
\end{center}
\caption{Comparison of average angular momenta calculated using
Eq.~(\protect\ref{lcor}) with a constant value (dashed curve) and
Eq.~(\protect\ref{r0ofe}) (solid curve) for $\rho_0$ to the values
calculated by IBMFUS (points) for the $^{16}$O+$^{154}$Sm system with the
quadrupole coupling strength $v_2=0.26$.  The parameters used were $r_{0} =
10.6$ fm, $\delta = 1.22$ fm, $V_{B0} = 59$ MeV, and $W = 2.3$ MeV.}
\label{fig7}
\end{figure}

The averaged radius of Eq.~\ref{r0ofe} can now be used in Eq.~(\ref{lcor})
in the place of $r_0$ to obtain an improved expression for average angular
momentum,
\begin{eqnarray}
\label{lcordef}
\langle\ell\rangle &=& \frac{\mu \rho_0^4(E) }{\hbar^2 E\sigma(E)} \int_0^E dE'
 \, \frac{E' \sigma(E')}{\rho_0^2(E')} \, \left\{
 \left[\frac{2\mu \rho_0^2(E)}{\hbar^2}\,(E-E')+ \frac{1}{4} \right]^{-1/2}
 \left[1-\frac{7}{\alpha \rho_0^2(E)} (E-E')\right] \right. \nonumber
\\[0.25cm]
& &\hspace{2.25cm} \left. +{4 \over \alpha} \left( {1 \over \rho_0^2(E')} - {1
\over \rho_0^2(E)} \right)
   \left( \left[\frac{2\mu \rho_0^2(E)}{\hbar^2}\,(E-E')+
       \frac{1}{4} \right]^{1/2} -{1\over 2} \right) \right\} \,.
\end{eqnarray}
Note that when $\rho_0(E)= r_0$, the last term vanishes and the above
expression reduces to Eq.~(\ref{lcor}).
The average angular momentum extracted from the fusion ``data" using
Eq.~(\ref{lcordef}) is compared to the $\langle \ell \rangle$ values
obtained from the same IBMFUS calculation in Fig.~\ref{fig7}.
The agreement is very good at all energies. In contrast, the constant radius
results underestimate $\langle \ell \rangle$ by about one $\hbar$.

\begin{figure}
\begin{center}
\epsfxsize=6.0in
\epsfbox{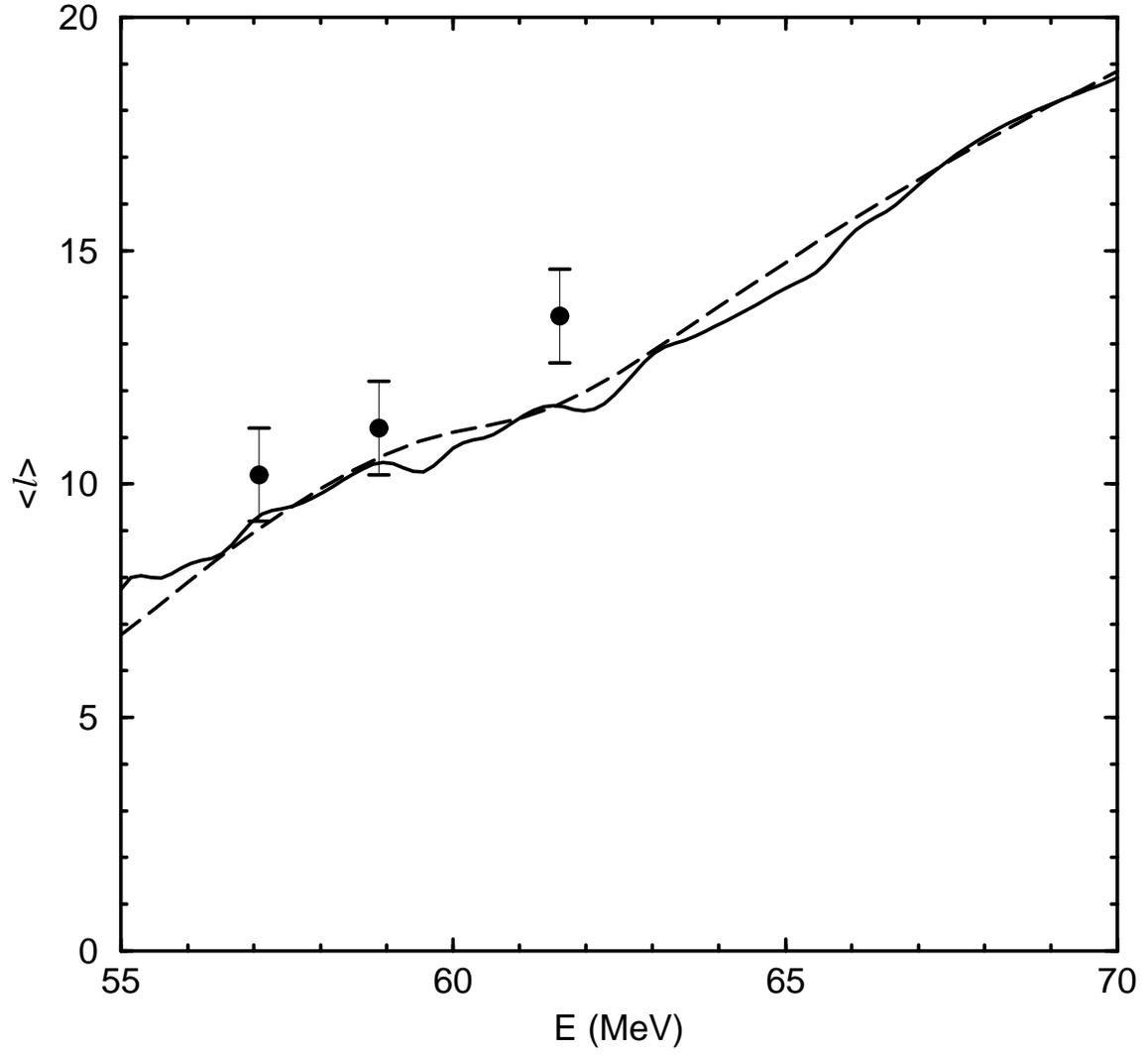}
\end{center}
\caption{Average angular momenta data\protect\cite{wei91} for the system
$^{16}$O+$^{154}$Sm.  The solid curve shows the results using
Eq.~(\protect\ref{lcordef}) and dashed curve those from
IBMFUS\protect\cite{bal94b}}
\label{fig8}
\end{figure}

To demonstrate the utility of this method in extracting $\langle \ell \rangle$,
we apply it to the $^{16}$O+$^{154}$Sm system for which quality cross section
data exist \cite{wei91}. In Fig.~\ref{fig8}, we compare the $\langle \ell
\rangle$ values obtained from Eq.~(\ref{lcordef}) and IBMFUS with the
experimental data \cite{bie93}. The two calculations are consistent with each
other but slightly underpredict the experimental values.


\section{Conclusions}

We have described two major reasons for the energy dependence of the
effective radius.  The effects of the centrifugal barriers are described by
using an improved approximation for the penetration probability.  This
also leads to a better relation between the cross section and the average
angular momentum.  The effects of target deformation are described
with a simple model which reproduces the features of the effective radius.
Finally, we present a phenomenological expression for the position of the
$s$-wave barrier as a function of energy and show how it can be used
in extracting the average angular momentum from the fusion cross section.
Comparison of this method with numerical calculations shows that it predicts
the average angular momentum from the fusion data reliably, and hence it
can be used as a consistency check in cases where quality data are available
for both quantities.


\section*{Acknowledgments}

This research was supported in part by the National Science Foundation
Grants No. PHY-9314131 and INT-9315876, and in part by the Australian
Research Council and by an exchange grant from the Department of
Industry, Science and Technology of Australia.


\appendix
\section*{The Validity of the Expansion of the Penetrability}

In this appendix, we discuss the validity
of Eq.~(\ref{tlcor}) which approximates the transmission probability as a
power series in $\Lambda=\ell(\ell+1)$.  We will do this
by using the linearized forms of the WKB penetration
integrals.  The penetration probability of an $\ell$-wave through a
one-dimensional barrier is given by
\begin{equation}
\label{a1}
T_{\ell}(E) = \left[ 1 + \exp \left( 2S_{\ell}(E) \right)\right]^{-1} ,
\end{equation}
where the WKB penetration integral is
\begin{equation}
S_\ell(E) = \sqrt{2\mu \over \hbar^2} \int_{r_{1\ell}}^{r_{2\ell}} dr
\left[ V_0(r) + {\hbar^2 \ell(\ell+1) \over 2\mu r^2} - E \right]^{1/2} .
\end{equation}
Using Abelian integrals, one can show that for energies below the barrier
\begin{equation}
\label{a2}
\int_E^{V_{B\ell}} dE' \, { S_{\ell}(E') \over \sqrt{E'-E}} = {\pi \over 2}
\sqrt{2\mu  \over \hbar^2} \int_{r_{1\ell}}^{r_{2\ell}} dr \left[ V_0(r) +
{\hbar^2
\ell(\ell+1) \over 2\mu r^2} - E \right] ,
\end{equation}
where $V_{B\ell}$ is the height of the $\ell$-wave potential, $V_0(r)$ is the
$s$-wave barrier, and $r_{1\ell}$, $r_{2\ell}$ are the turning points of the
$\ell$-wave barrier for energy $E$. In Refs. \cite{bal83,col78}, the energy
derivative of Eq.~(\ref{a2}),
\begin{equation}
\label{a3}
\int_E^{V_{B\ell}} dE' \, { \partial S_{\ell}(E')/\partial E' \over
\sqrt{E'-E}}
= - {\pi \over 2} \sqrt{2\mu \over \hbar^2} \; (r_{2\ell} - r_{1\ell}) \,,
\end{equation}
 was used to find the barrier thickness.
One can also take the
derivative of Eq.~(\ref{a2}) with respect to $\Lambda$ to obtain
another useful identity \cite{col78},
\begin{equation}
\label{a4}
\int_E^{V_{B\ell}} dE' \, { \partial S_{\ell}(E')/\partial \Lambda \over
\sqrt{E'-E}}
= - {\pi \over 2} \sqrt{\hbar^2 \over 2\mu} \left({1 \over r_{2\ell}} - {1
\over r_{1\ell}}\right).
\end{equation}

The last two equations can be used to check the consistency of an
expansion of the form
\begin{equation}
\label{a5}
S_{\ell} (E) = S_0 (E- e_1 \Lambda - e_2 \Lambda^2 - ...) \,,
\end{equation}
where $e_1,e_2,...$ are independent of the energy. Substituting the
relationship between the derivatives,
\begin{equation}
\label{a6}
{\partial S_{\ell} (E) \over \partial \Lambda}  = - (e_1 + 2 e_2 \Lambda + ...)
{\partial S_{\ell} (E) \over \partial E} \,,
\end{equation}
into Eqs. (\ref{a3}) and (\ref{a4}), we find that the assumption in
Eq.~(\ref{a5}) is consistent if
\begin{equation}
\label{a7}
(e_1 + 2 e_2 \Lambda + ...) = {\hbar^2 \over 2\mu r_{1\ell}r_{2\ell}} \,.
\end{equation}
The right hand side of this equation has a slight dependence on energy
(through the energy-dependence of the turning points) whereas the left hand
side is independent of energy in our approximation.
Our analysis in Section \ref{sec-ldep} is equivalent to approximating the
right hand side of Eq.~(\ref{a7}) by a constant, {\it i.e.}
\begin{equation}
\label{a8}
r_{1\ell}r_{2\ell} \sim r_{\ell}^2 \,,
\end{equation}
where $r_{\ell}$ is the position of the peak of the $\ell$-wave barrier.
This is a very good approximation at energies near the barrier height, but
the error increases as the energy gets lower. However, for the $^{16}$O
$+^{154}$Sm system, even at 7 MeV below the barrier (lower than the fusion
cross section has yet been measured) the error in this approximation is
less than 4 \% for the bare potential used in Ref.~\cite{bal94a}.  Upon
setting $\ell=0$ in Eq.~(\ref{a8}), we recover
\begin{equation}
\label{a9}
e_1 = {\hbar^2 \over 2\mu r_0^2} \,,
\end{equation}
in agreement with Eq.~(\ref{tlcor}).  The higher-order terms in $\Lambda$ also
agree with the calculations in Section \ref{sec-ldep}, so we are justified
in expressing $T_\ell(E)$ as a power series in $\Lambda$.

It should be noted that although $\Lambda$ is not a small parameter, there
is a natural cutoff $\Lambda_{cr}$ in this parameter given by the condition
\begin{equation}
\frac{\Lambda_{cr} \hbar^2}{2\mu R^2} \simeq E \,.
\end{equation}
For values of $\Lambda$ greater than this, fusion will not occur.
Due to this cutoff, the second term in Eq.~(\ref{tlcor}) will always be
larger than the third term.  Therefore, the additional term can be
considered a correction.


{}


\newpage

\end{document}